\documentclass[preprint]{elsarticle}

\usepackage{lineno,hyperref}
\graphicspath{{Figures/}}

\usepackage{siunitx}

\usepackage{subfig}
\usepackage{multirow}
\modulolinenumbers[5]
\usepackage{geometry} 
\usepackage{multicol}
\usepackage{tabularx,ragged2e,booktabs,caption}
\biboptions{numbers,sort&compress}

\journal{NIM-A}

\begin{document}

\begin{frontmatter}

\title{Double-crystal measurements at the CERN SPS}

\author[CERN_address,ICL_address,INFN_address2]{W.~Scandale}
\cortext[correspondingauthor]{Corresponding authors: walter.scandale@cern.ch (W. Scandale) }
\author[CERN_address]{G.~Arduini}
\author[CERN_address]{F.~Cerutti}
\author[CERN_address,INFN_address4]{M.~D'Andrea}
\author[CERN_address]{L.S.~Esposito}
\author[CERN_address,ICL_address]{M.~Garattini}
\author[CERN_address]{S.~Gilardoni}
\author[CERN_address,manchester_address]{D.~Mirarchi}
\author[CERN_address]{S.~Montesano}
\author[CERN_address,PSUD_address,TSUNK_address]{A.~Natochii}
\author[CERN_address]{S.~Redaelli}
\author[CERN_address]{R.~Rossi}
\author[CERN_address,JINR_address]{G.I.~Smirnov}
\author[PSUD_address]{L.~Burmistrov}
\author[PSUD_address]{S.~Dubos}
\author[PSUD_address]{V.~Puill}
\author[PSUD_address]{A.~Stocchi}
\author[CERN_address,INFN_address2]{F.~Addesa}
\author[INFN_address1]{F.~Murtas}
\author[INFN_address3]{F.~Galluccio}
\author[JINR_address]{A.D.~Kovalenko}
\author[JINR_address]{A.M.~Taratin}
\author[PNPI_address]{A.S.~Denisov}
\author[PNPI_address]{Yu.A.~Gavrikov} 
\author[PNPI_address]{Yu.M.~Ivanov}
\author[PNPI_address]{L.P.~Lapina} 
\author[PNPI_address]{L.G.~Malyarenko}
\author[PNPI_address]{V.V.~Skorobogatov}
\author[IHEP_address]{A.G.~Afonin}
\author[IHEP_address]{Yu.A.~Chesnokov}
\author[IHEP_address]{A.A.~Durum}
\author[IHEP_address]{V.A.~Maisheev}
\author[IHEP_address]{Yu.E.~Sandomirskiy}
\author[IHEP_address]{A.A.~Yanovich}
\author[ICL_address]{J.~Borg}
\author[ICL_address]{T.~James}
\author[ICL_address]{G.~Hall}
\author[ICL_address]{M.~Pesaresi}

\address[CERN_address]{The European Organization for Nuclear Research (CERN), CH-1211 Geneva 23, Switzerland}
\address[ICL_address]{Blackett Laboratory, Imperial College, London SW7 2AZ, United Kingdom}
\address[PSUD_address]{Laboratory of Physics of 2 Infinities Irene Joliot-Curie (IJCLab), Orsay, France}
\address[TSUNK_address]{On leave from Taras Shevchenko National University of Kyiv (TSNUK), 60 Volodymyrska Street, 01033 Kyiv, Ukraine}
\address[INFN_address1]{INFN, Laboratori Nazionali di Frascati, Via Fermi, 40 00044 Frascati (Roma), Italy}
\address[INFN_address2]{INFN Sezione di Roma, Piazzale Aldo Moro 2, 00185 Rome, Italy} 
\address[INFN_address3]{INFN Sezione di Napoli, Complesso Universitario di Monte Sant'Angelo, Via Cintia, 80126 Napoli, Italy}
\address[INFN_address4]{Dipartimento di Fisica e Astronomia "Galileo Galilei", Università degli Studi di Padova, Via Marzolo 8, I-35131 Padova, Italy}
\address[JINR_address]{Joint Institute for Nuclear Research (JINR), Joliot-Curie 6, 141980 Dubna, Russia}
\address[PNPI_address]{Petersburg Nuclear Physics Institute in National Research Centre "Kurchatov Institute", 188300 Gatchina, Russia}
\address[IHEP_address]{NRC Kurchatov Institute - IHEP, 142281 Protvino, 142281 Russia}
\address[manchester_address]{Presently also at The University of Manchester, Manchester M13 9PL, United Kingdom}

\begin{abstract}
The UA9 setup, installed in the Super Proton Synchrotron (SPS) at CERN, was exploited for a proof of principle of the \textit{double-crystal scenario}, proposed to measure the electric and the magnetic moments of short-lived baryons in a high-energy hadron collider, such as the Large Hadron Collider (LHC). Linear and angular actuators were used to position the crystals and establish the required beam configuration. Timepix detectors and high-sensitivity Beam Loss Monitors were exploited to observe the deflected beams. Linear and angular scans allowed exploring the particle interactions with the two crystals and recording their efficiency. The measured values of the beam trajectories, profiles and of the channeling efficiency agree with the results of a Monte-Carlo simulation.
\end{abstract}

\end{frontmatter}
                                            


\section{Introduction}

Planar channeling in bent crystals is a powerful tool often exploited for beam manipulations in particle accelerators~\cite{CHANNELING_2}.
The angular acceptance for channeling is determined by the critical angle $\theta_{\rm c} = \sqrt{2U_{\rm max}/pv}$, where $U_{\rm max}$ is the potential well between two neighbouring crystalline planes ($\rm\sim20$~eV for Si (110)), $p$ and $v$ are the particle momentum and velocity respectively.
The UA9 experimental ﬁndings on the interaction of high-energy particles with bent crystals are reported in~\cite{scandale2019channeling}, where the studies performed over more than a decade at the CERN accelerator complex by the UA9 Collaboration are described. The exploitation of bent crystals in the NA48 experiment, for a direct proof of CP violation with secondary $K_S$, $K_L$  meson beams in the CERN North Area, is reported in~\cite{NA48}. The bent crystal based measurement of the magnetic moment of the $\Sigma^+$ hyperon in the E761 apparatus at one of the Tevatron extraction lines at FNAL is described in~\cite{E761_1,fnal_MDM}. 
Beam extraction tests assisted by bent crystals performed at the SPS are discussed in~\cite{UA9_RD22_5,UA9_RD22_8}. Similar tests performed at the Tevatron at FNAL are reported in~\cite{Carrigan1994,carrigan1996_1,Carrigan2002beam}. Studies on beam-halo collimation based on bent crystals, performed in RHIC at NBNL, in the Tevatron, in the SPS and in LHC are reported in~\cite{flillerRHIC1998,UA9_3,hilumi-LHCC-LARP2011,Mokhov2010crystal,Scandale2010SPS1,Mirarchi2011,rossi2018experimental}.

Recently, a \textit{double-crystal scenario} has been proposed to measure the magnetic moment of short-lived baryons in a high-energy hadron collider such as the LHC \cite{DOUBLECHANNELING3}. The basic layout consists of a sequence of two crystals interleaved with a solid target, installed in the vicinity of an experimental detector that should allow identifying the rare baryons and recording their polarization state from the reconstruction of the decay products. The first crystal should capture in channeling states part of the halo of the circulating beam and deflect it onto the internal target, where baryons could be produced. The second crystal should capture a fraction of the baryons, rotate their polarisation vector and simultaneously deflect them into the detector. If implemented in LHC, the first crystal could be similar to those used as primary crystal-collimator in the LHC beam collimation studies~\cite{EXTRA7}. The second crystal should have a larger deflection angle in order to induce a large enough rotation of the polarization vector for an accurate detection of the magnetic dipole moment, as discussed in~\cite{EXTRA5,EXTRA1,EXTRA2,EXTRA3,EXTRA4,mirarchi2020layouts}. 

Investigating such a sophisticated beam manipulation at lower beam energy in a circular accelerator with warm iron magnets is the best way to assess its feasibility and evaluate its performance in safer conditions than in a superconducting collider. A simplified version of the double-crystal configuration was recently implemented in the SPS, for a proof-of-principle. The setup results from successive modifications of the UA9 layout~\cite{EXTRA9}, which had originally been conceived to test the crystal-assisted collimation process~\cite{UA9_12}. The initial double-crystal layout and the results recorded in the first SPS test are discussed in~\cite{MyPHD}. 
Operational aspects of the upgraded layout and recent results from the data collected for the ﬁrst time in a circular accelerator are reported hereafter, together with their comparison with computer simulations.

\section{The double-crystal setup in the CERN-SPS}

\begin{figure}[t]
  \centering
  \includegraphics[width=1\linewidth]{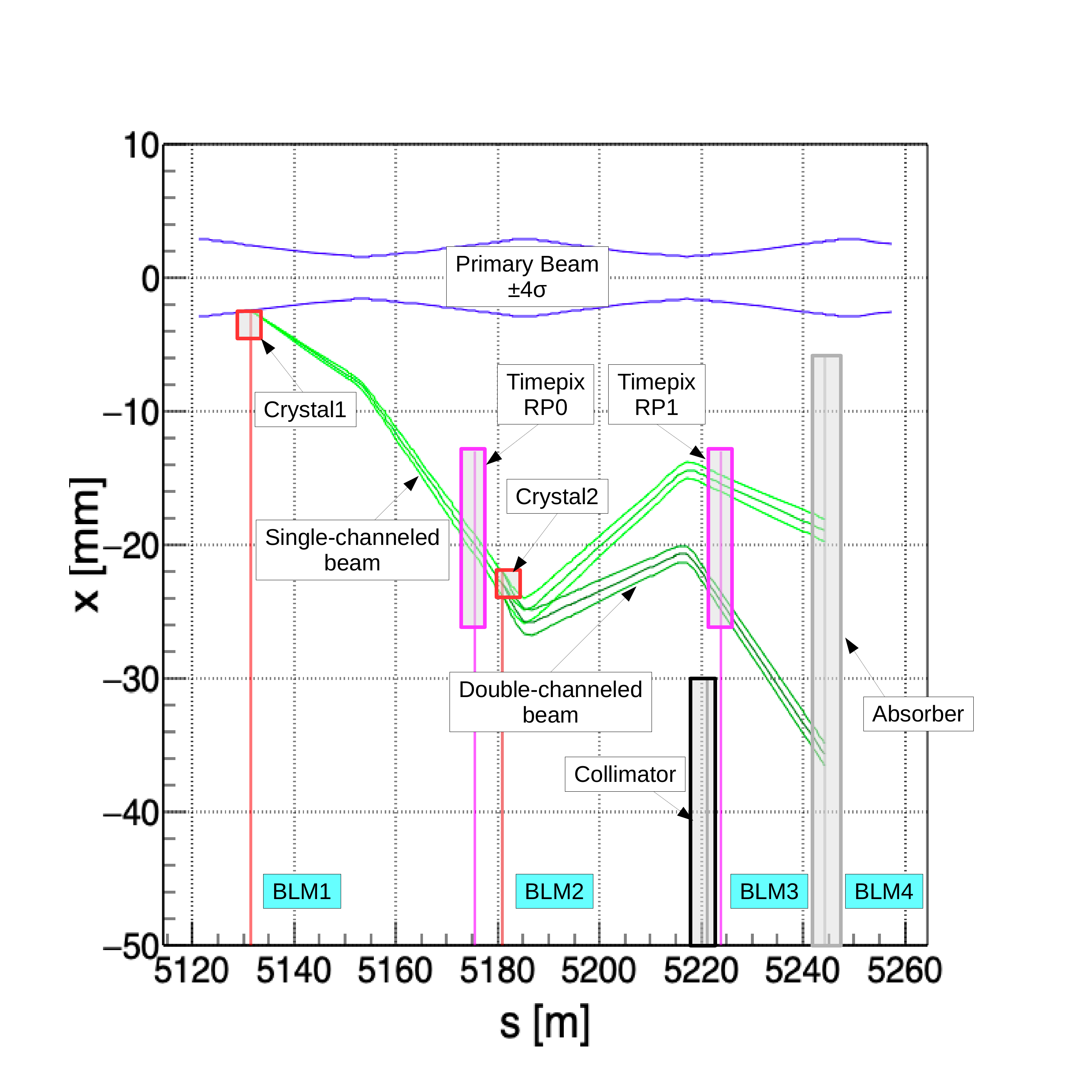}
    \caption{Schematic view of the UA9 double-crystal setup operating on a stored proton beam of 270 GeV energy, with $\epsilon_{H} =~5~$nm$\cdot$rad horizontal emittance. The beam envelope at 4$\sigma$ is shown in blue, the two crystals in red, the two Roman pots equiped with Timepix sensors in pink, the collimator for beam-based alignement in black and the secondary absorber in gray. The single-channeled beam envelope is drawn in light green and the double-channeled one in dark green. BLMs positions are indicated in light blue. The horizontal axis gives the distance from the conventional origin of the SPS circle and the vertical axis represents the horizontal distance from the center-line of the beam pipe. \\}
\label{fig:fig1}
\end{figure}

\begin{figure}[t]
    \centering\includegraphics[width=0.85\linewidth]{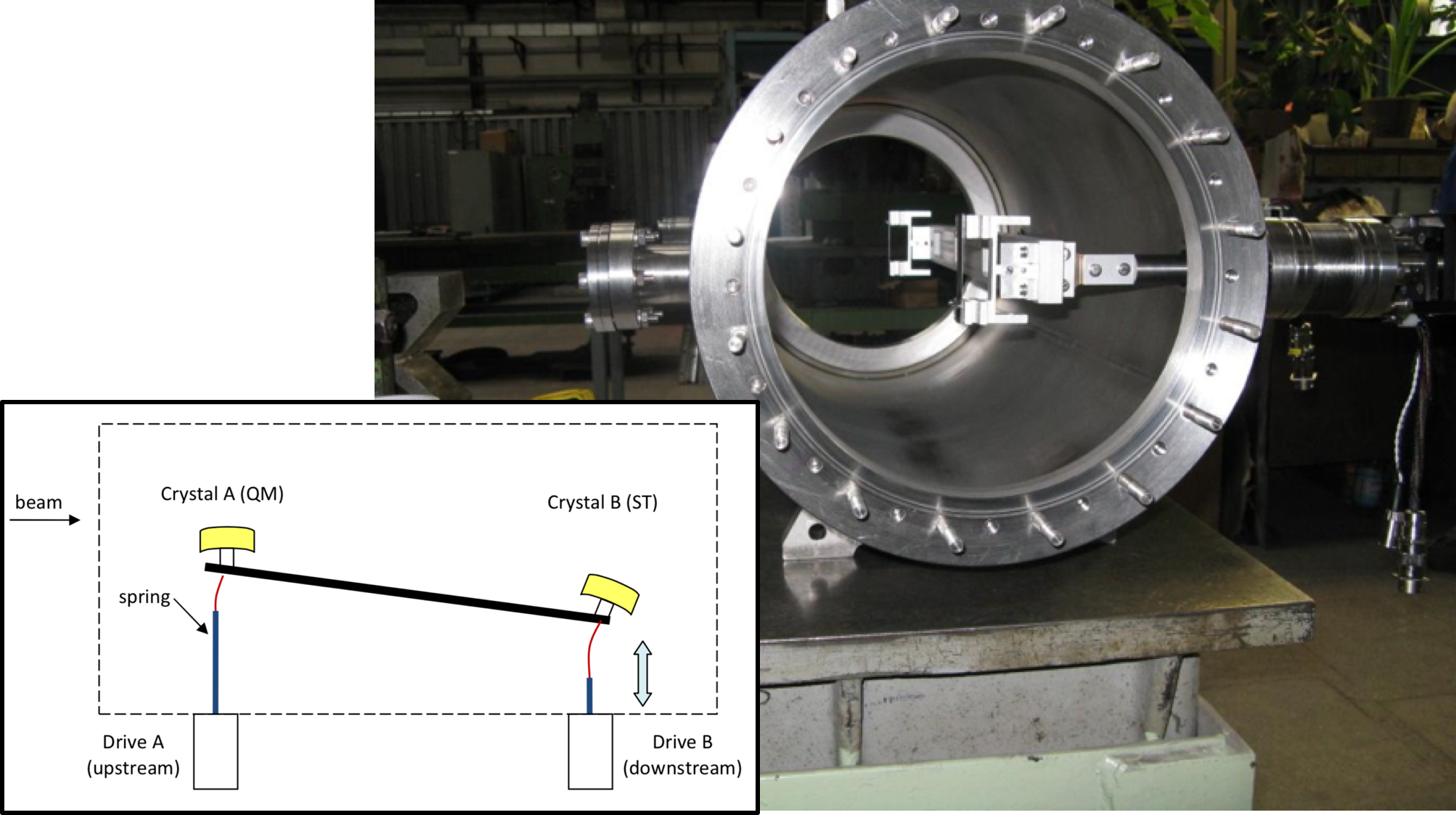}
    \caption{Mechanical actuator of the UA9 crystals. Each vacuum tank can house two crystals independently adjustable for channeling. Two linear actuators linked by a bar are visible on the right. One bar can position the crystal at the edge of the circulating beam and the other, slightly retracted, can change the crystal angle. In the insert, the scheme of principle of the angular actuator is shown.  \\}
\label{fig:fig2}
\end{figure}

The SPS is the CERN syncrotron used for fixed target physics in the North Area and as injector for the LHC. It is made of elements operating at room temperature, in which proton and heavy ion beams can be either accelerated up to 450 GeV per charge or stored for several hours at a plateau of up to 270 GeV per charge. Flexible operational conditions and resilience to beam loss make the SPS the ideal accelerator to investigate unusual beam manipulations.
The double-crystal configuration implemented in the SPS is schematically shown in Figure~\ref{fig:fig1}. The blue lines identify the circulating beam envelope at 4$\sigma$. When properly positioned and oriented, the upstream Crystal1 should channel and deflect the halo particles impinging within the critical angular range $\pm\theta_{\rm c}$, producing a single-channeled beam (light green lines). The downstream Crystal2 can be moved along the transverse horizontal direction to intercept the single-channeled particles and can channel them again, producing a double-channeled beam (green lines). Both channeled beams can be directed onto the tungsten absorber, 60 cm long (grey box), and dumped. Two collimator jaws can horizontally enclose the circulating particles to define sharp edges of the beam envelope for the beam-based alignment of the movable devices. 
Each component of the UA9 setup is followed by a high-sensitivity Beam Loss Monitor (BLM)~\cite{EXTRA8}, outside the beam pipe, to record secondary particle cascades produced by particle interactions during operation of the movable devices.
The movable Roman pots RP0 and RP1 (pink boxes) contain Timepix detectors~\cite{UA9_36} to measure the deflected beam profiles. 
The Timepix is a hybrid pixel detector developed by the CERN-based European Medipix2 collaboration~\cite{medipix}. It consists of a CMOS readout circuit (Timepix) bump-bonded to a 300~\SI{}{\micro\meter} thick silicon pixel detector with a matrix of 256$\times$256 square pixels of 55~\SI{}{\micro\meter} pitch, for a total area of 1.4 $\times$ 1.4 cm$^2$. Each cell in the matrix contains a charge-sensitive preamplifier, a single threshold discriminator, time based logic and a 14-bit pseudo-random counter with overflow control logic~\cite{ESPOSITO2011}. In Figure~\ref{fig:fig2} one of the two UA9 angular actuators, often named \textit{goniometer}, is shown: it is made of two linear actuator distant 50 cm from each other linked by an aluminium bar to stabilise the mechanical structure. Each goniometer houses two crystals that can be independently positioned and oriented, during an angular scan. One of the linear actuators can position the corresponding crystal at the edge of the circulating beam and the other, slightly retracted, can change the crystal angle. The movable structure of the goniometer, shown in the insert of Figure~\ref{fig:fig2},  is  conceptually elastic and friction free, however, backlash, hysteresis and non-linearities, detrimental for alignment accuracy, are in practice unavoidable. Lengthy and delicate corrective actions, investigated and calibrated in the laboratory and in the SPS test runs, are introduced to control the crystal orientation to the level of 2-3~\SI{}{\micro\radian} RMS. They consist in imposing to reach a given motor positioning always through the same mechanical cycle, thus avoiding inversion of motion for fine tuning of crystal position and compensating the backlash at the end of the alignment cycle. Static orientation of the crystal performed without applying the corrective actions may result in a large shift from the required angle. Moreover, during dynamical scans of the crystal orientation performed at constant rotational speed (angular scans), the angle versus time correlation is substantially linear: indeed, the one-directional continuous motion guarantees the absence of backlash and residual non-linear angular variations smaller than 2-3~\SI{}{\micro\radian} RMS.  

\begin{figure}[t]
  \centering
  \includegraphics[width=1\linewidth]{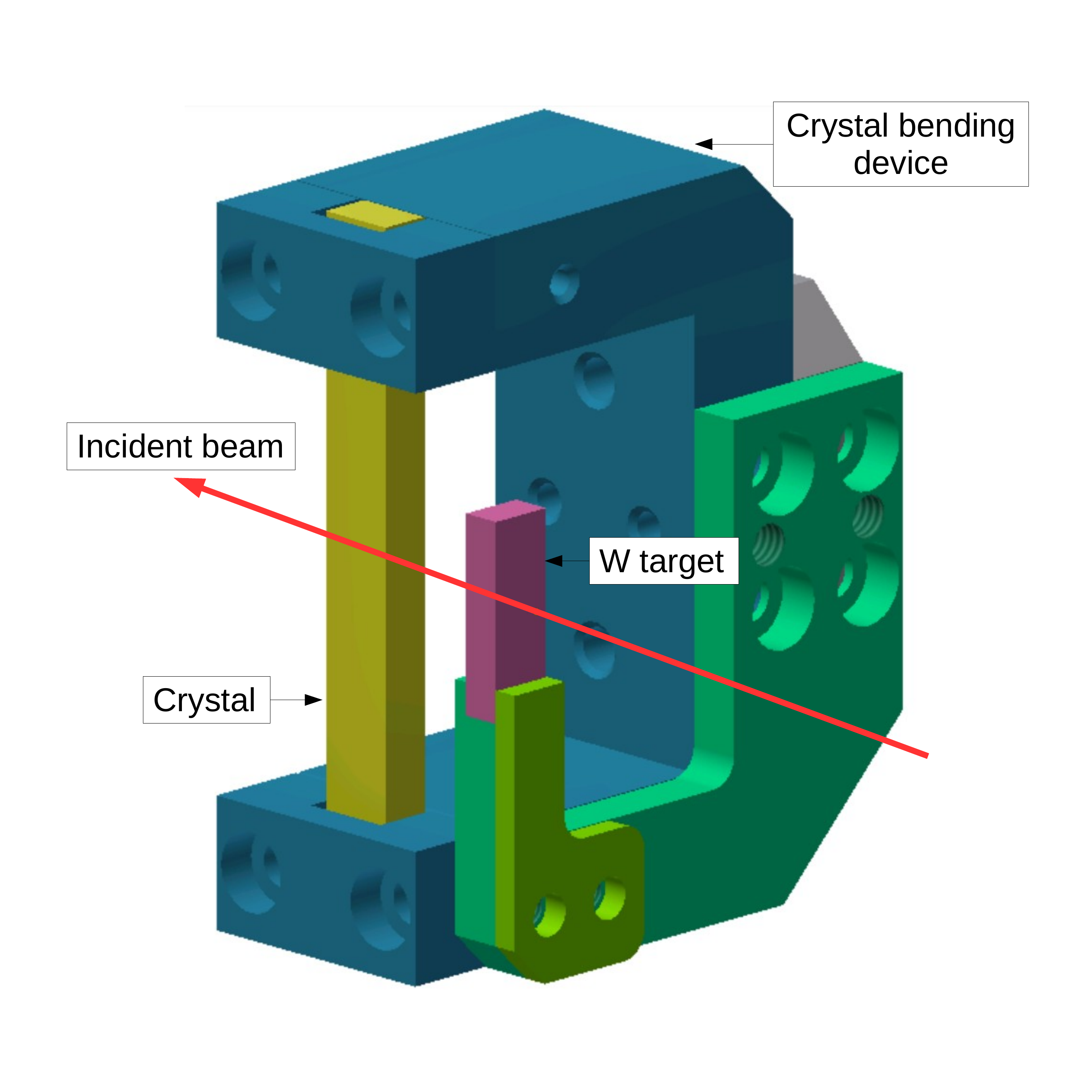}
    \caption{Schematic view of the Crystal2 with the target attached to its frame. The green support of the tungsten target, drawn in pink, is screwed to the Crystal2 holder, with the crystal and the target faces parallel to each other. The red arrow indicates the direction of the incident particles single-channeled by Crystal1.\\}
\label{fig:fig3}
\end{figure}

During data-taking runs, a single bunch of $\rm\sim 10^{11}$ protons was injected, accelerated and stored in the SPS at 270 GeV energy, with a typical value of the horizontal emittance $\epsilon_{H} =$~5~nm$\cdot$rad and with betatron tunes $Q_{H} = 20.13$, $Q_{V} = 20.18$. The characteristics of Crystal1 and Crystal2 are shown in Table~\ref{crystals}. A tungsten target was inserted in front of Crystal2 just before the very last SPS run, before the CERN accelerator Long Shutdown 2~\cite{FixedTargetH8}.  The target was mounted on the crystal bending frame according to the scheme of Figure~\ref{fig:fig3}. The target dimensions are indicated in the third column of Table~\ref{crystals}, together with the RMS deflection imparted to 270 GeV protons by Multiple Coulomb Scattering (MCS). The betatron functions of interest are shown in Table~\ref{betatron}. The data discussed in Sections 3 to 6 were collected without the W-T target, those in Section 7 were recorded with the target in place. The latter ones provide an insight to the additional background that should be rejected for the optimal exploitation of the physics reach in the double crystal scenario.
\begin{table}
   \centering
   \caption{Characteristics of the crystals and of the target}
   \resizebox{\columnwidth}{!}{%
   \begin{tabular}{lccc}
       \toprule
     
             & \textbf{Crystal1} & \textbf{Crystal2} & \textbf{Target} \\
	\midrule     
       	Name	&\textbf{ACP78} & \textbf{ACP75} & \textbf{W-T}\\

          Deflection [\SI{}{\micro\radian}] & 301       & 197   &46*  \\ 
          Length [\SI{}{mm}] & 4.0           & 6.0       & 3.0\\ 
          Width [\SI{}{mm}]  & 1.5            & 4.0    & 5.0\\ 
          Torsion [\SI{}{\micro\radian/mm}]  & 3.0            & 4.0    \\ 
          $\theta_{\rm c}$ [\SI{}{\micro\radian}] for 270 GeV $p^{+}$ & 12.19    &12.25  \\ 
       \bottomrule
         
   \end{tabular}
   }
   \label{crystals}
   *RMS angle imparted to 270 GeV protons by MCS
\end{table}

\begin{table}
   \centering
   \caption{Betatron functions of the double-crystal layout}
   \resizebox{\columnwidth}{!}{%
   \begin{tabular}{lcccccc}
  
       \toprule
     
         Element & s [\SI{}{m}] & $\beta_{x}$ [\SI{}{m}] & $\beta'_{x}$ & $D_{x}$ [\SI{}{m}] & $D'_{x}$ & $\Delta \Phi_{x}$ \\
	\midrule     

          Crystal1      & 5121  & 74.68   &1.49 &-0.45  &-0.0005    &0  \\ 
          RP0           & 5175  & 76.04   &-1.51      & -0.851 &-0.0174      & 0.152\\ 
          Crystal2      & 5181  & 93.59   &-1.74      & -0.946 &-0.0174      & 0.163\\ 
          Collimator    & 5222  & 36.74   &-0.761      & -0.419 &-0.0027      & 0.285\\ 
          RP1           & 5224  & 40.17   &-0.852      & -0.415 &-0.0027      & 0.293\\ 
        Absorber        & 5244  & 92.90   &-1.730      & -0.358 &-0.0027      & 0.348\\ 
       \bottomrule
         
   \end{tabular}
   }
   \label{betatron}
\end{table}

\section{Optimizing the double-crystal operation}

\begin{figure}[t]
    \centering\includegraphics[width=1\linewidth]{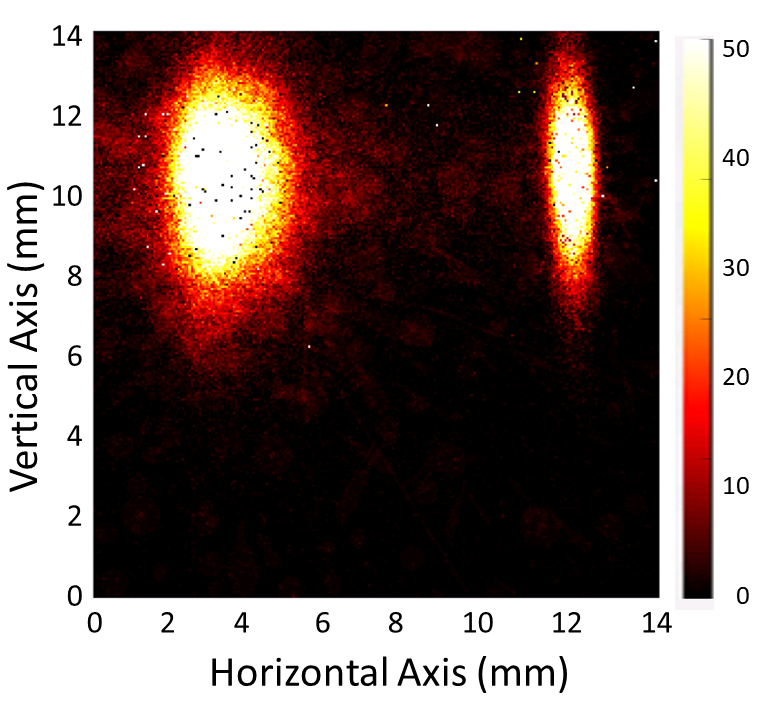}
    \caption{A frame of Timepix in RP1 recorded during the double crystal experimental run. The spot on the left is the image of the single-deflected beam, while the spot on the right shows the double-deflected one.\\}
\label{fig:fig4}
\end{figure}

\begin{figure}[t]
    \centering
    \includegraphics[width=1\linewidth]{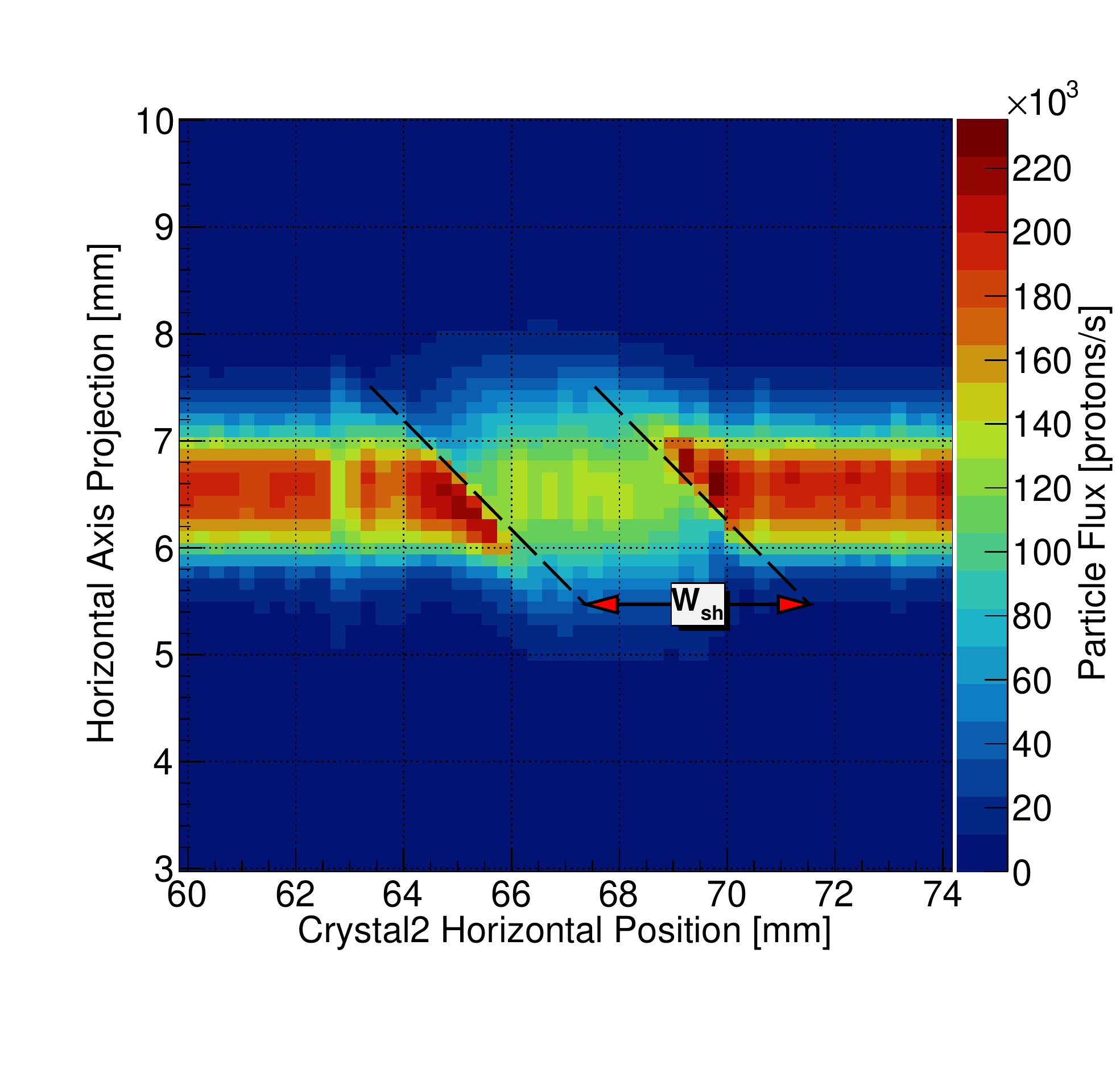}
    \caption{Horizontal projection of the single-deflected beam profile as a function of Crystal2 horizontal position, as detected by Timepix in RP1 during a linear scan. The profile is broadened when Crystal2 interacts with the single-deflected beamlet. The dashed line in black defines the interaction area. $W_{sh}$ is equal to Crystal2 width.  The color code scale indicates the particle flux.\\}
\label{fig:fig5}
\end{figure}

\begin{figure}[t]
    \centering\includegraphics[width=1\linewidth]{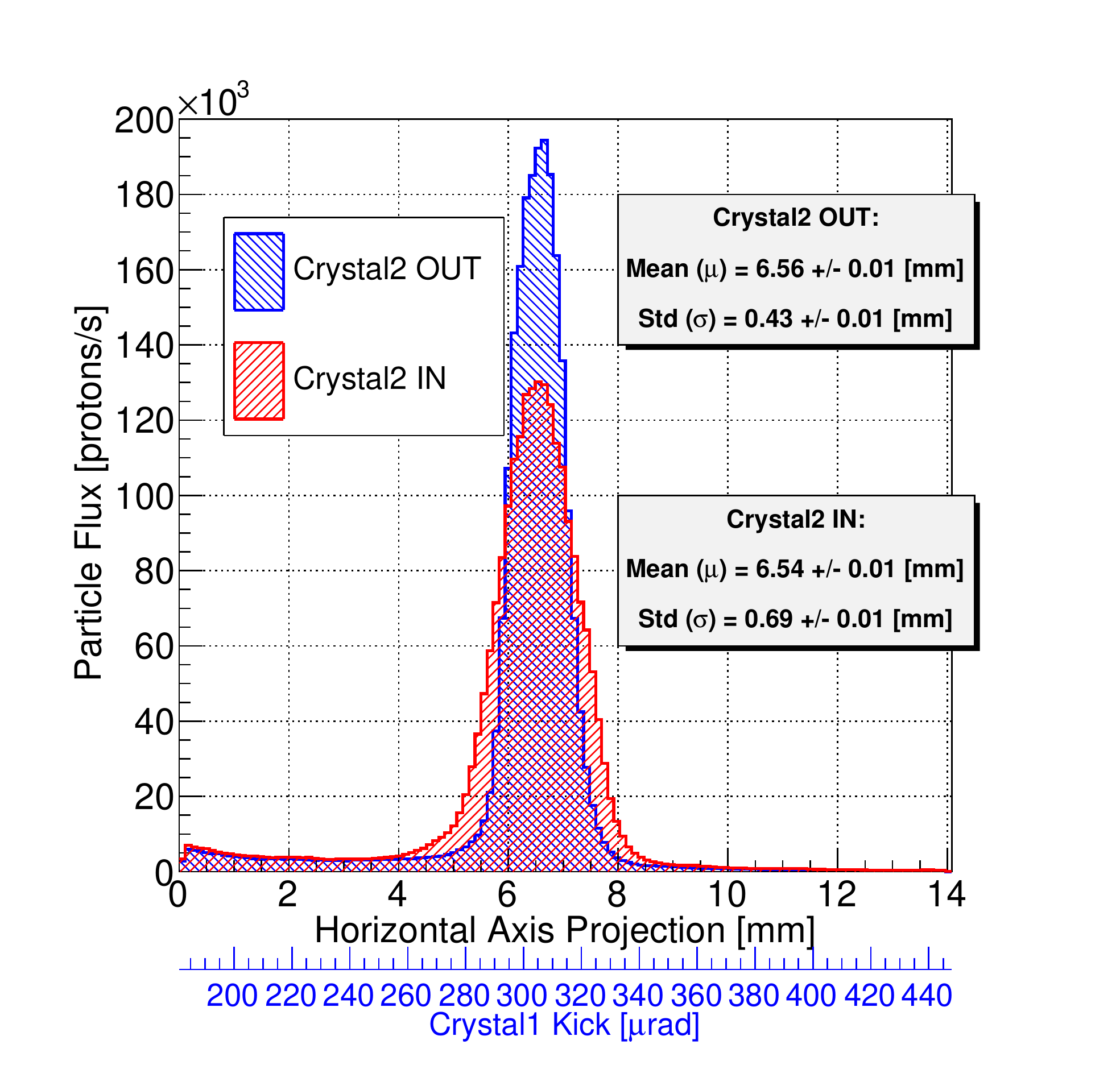}
    \caption{Horizontal single-channeled beam profile measured by the Timepix in RP1. "Crystal2 IN": Crystal2 is centered onto the single-deflected beam in amorphous orientation; "Crystal2 OUT": Crystal2 is retracted from the beam.\\}
\label{fig:fig6}
\end{figure}
 
Setting-up a single-channeled beam in the SPS is a well-established operation that can be reliably achieved with the UA9 devices~\cite{firstcollimation2010}. In order to capture halo particles in channeling states Crystal1 is positioned at the circulating beam periphery  and its orientation is slowly rotated by the angular actuator supporting the crystal. Its optimal orientation is obtained by identifying the angle at which the particle loss rate in the BLM1 downstream is minimum. 

In double channeling operation mode, intercepting the single-deﬂected beam with Crystal2 and ﬁnding its optimal orientation for channeling requires a more complex procedure and more sensitive detectors due to the much lower ﬂux of particles being deflected by Crystal1. The Timepix detector in the RP1, inserted into the vacuum pipe during the alignment process, can considerably ease the search for the optimal orientation by continuously providing the image of both the double- and single-channeled beam spots. An example of frame recorded by the Timepix in RP1 is shown in Figure~\ref{fig:fig4}, where the single-deflected spot is on the left side and the double-deflected one is on the right side. To improve the display and the subsequent data analysis, in each frame, the electronic noise is compensated and the distorted signal of the pixels irradiated by a too large local flux is filtered and smoothed taking into account the average response of the surrounding pixels.

A linear scan of Crystal2 from its garage position towards the circulating beam edge allows identifying its optimal superposition with the single-deflected beam. Keeping the crystal in the amorphous orientation during the scan maximizes the interaction rate and induces a better detectable shadow in the beam image recorded by the Timepix in RP1 during the superposition. Such a procedure is well illustrated in Figure~\ref{fig:fig5}. The crystal moves at a constant speed of 50~\SI{}{\micro\meter}/s. The horizontal axis indicates the instantaneous Crystal2 distance from its garage position. The vertical axis shows the horizontal projection of the corresponding Timepix image recorded and averaged progressively every 0.5 s. A color code quantifies the local flux integrated over each pixel column. The beam spot projection is almost constant for most of the scan duration. When moving from 64.4 mm to 70.4 mm, the crystal position is partially or totally superimposed to the deflected beam trajectory and the recorded profile is broadened because of MCS. The enlarged region appears inclined because of the finite sizes of the beam and of the crystal. Moreover, as expected, its horizontal extension $W_{\rm sh} = 4.15 \pm 0.14$~mm is equal to the Crystal2 width, indicated in Table~\ref{crystals}.
The maximum superposition of Crystal2 with the single-deflected beam is obtained at $\sim$67 mm. In Figure~\ref{fig:fig6}, the horizontal profile of the single-channeled beam is shown as a function of the horizontal distance from the inner border of the Timepix sensor in RP1, on the side of the circulating beam. The blue scale under the horizontal axis translates an abscissa value into the corresponding angle at Crystal1, through the appropriate transport matrix. The blue curve shows the unperturbed profile, while the red one shows the perturbed profile, when Crystal2 is inserted in amorphous orientation at 67 mm position. The mean and the RMS values of the blue distribution are 6.56$\pm0.01$ mm and 0.43$\pm0.01$ mm, those of the red one 6.54$\pm0.01$ mm and 0.69$\pm0.01$ mm, respectively (see also the inserts in Figure~\ref{fig:fig6}). The profile mean value, practically unaffected by Crystal2 insertion,  corresponds to the deflection angle $\theta_{def}$~=~305$\pm2~\SI{}{\micro\radian}$, very close to the nominal bending angle of Crystal1 (see Table~\ref{crystals}). The RMS width of the red and blue curves differ because of the MCS occurring in Crystal2. Simulations provide RMS values of 0.56 mm for Crystal2 OUT and 0.79 mm for Crystal2 IN, in reasonable agreement with measurements.
 
\begin{figure}[t]
    \centering\includegraphics[width=1\linewidth]{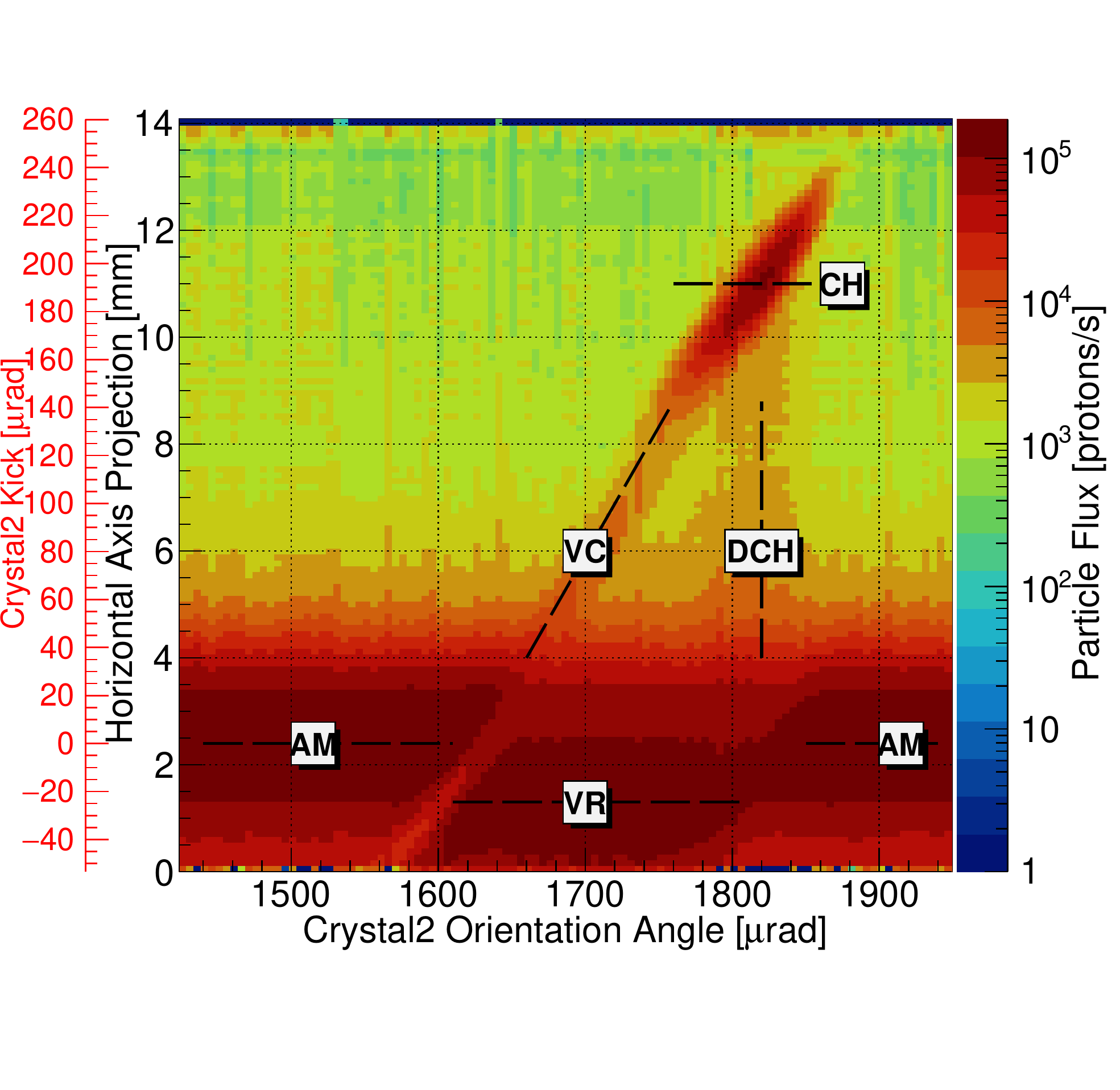}
    \caption{Timepix image projection on the horizontal axis as a function of Crystal2 orientation angle. The black dashed lines highlight the various interaction areas: CH (channeling), VR (volume reflection), AM (amorphous interactions), DCH (dechanneling) and VC (volume capture). The vertical scale in black refers to the horizontal beam profile recorded by the Timepix in RP1, integrated over 0.5 s. The vertical scale in red indicated the kick imparted by Crystal2 to obtain the horizontal particle displacements given in black. The color code  indicates the particle flux per pixel.\\}
\label{fig:fig7}
\end{figure}

Once Crystal2 is centered in the single-channeled beam at 67 mm, an angular scan should help finding the optimal orientation for channeling. The procedure consists in rotating the crystal at constant speed, typically of 2~\SI{}{\micro\radian}/s, while recording the signal of the Timepix in the RP1.  
Figure~\ref{fig:fig4} illustrates a typical way of plotting the results. 
The horizontal axis shows the Crystal2 angle and the vertical axis the horizontal projection of the corresponding Timepix frames progressively collected and averaged every 0.5 s. A color code quantifies the local flux.
Crystal2 is initially in amorphous orientation (AM) and the single-channeled beam is just slightly broadened because of the MCS. At around 1600~\SI{}{\micro\radian} the volume reflection (VR) starts to dominate for about 200~\SI{}{\micro\radian}. As expected, the width of the VR angular range is very close to the nominal bending angle of Crystal2 (Table~\ref{crystals}). The VR process is accompanied by volume capture (VC), its competing effect, whose probability is much smaller. The channeling process (CH) appears around 1800~\SI{}{\micro\radian} simultaneously with the lower-probability dechanneling effect (DCH). In the optimal CH orientation, at $1820~\SI{}{\micro\radian}$, the displacement of the double-deflected beam from the baricentre of the single-deflected beam is 8.5 mm. Such a value corresponds to a deflection $\theta_{def}=192.79\pm0.15~\SI{}{\micro\radian}$ imparted by Crystal2, as also shown by the red scale to the left of Figure~\ref{fig:fig7} that translates displacements at the RP1 into angles at Crystal2, through the betatron transfer matrices. As expected, $\theta_{def}$ is in good agreement with the crystal bending angle in Table~\ref{crystals}. 

The plot in Figure~\ref{fig:fig7} is very similar to those produced at the external beam-line H8 in the SPS North Area, where crystals are usually tested and characterized, before their installation in the SPS ring~\cite{CHANNELING_14}. 
The crystal alignment in the double-crystal setup could also be obtained using angular scans in which the loss data are collected by the high-sensitivity BLMs. However, the resulting procedure might become less straightforward  because of the low level of induced losses, especially downstream Crystal2~\cite{Montesano_2018}. 

In the double-channeling configuration implemented in the SPS, an undesired effect contributes to increase the fraction of particles traversing Crystal2 in AM orientation, thus decreasing the channeling efficiency. At the exit of Crystal1, the RMS  
divergence of the channeled beam is $\delta_{\rm RMS}\theta^{\rm CR1}_{\rm def} = \theta^{\rm CR1}_{\rm c} = 12.19~\SI{}{\micro\radian}$. At the entrance of Crystal2, it becomes $\delta_{\rm RMS}\theta^{\rm CR2}_{\rm def} = 18.9~\SI{}{\micro\radian}$, as a result of the quadratic composition of the betatronic transport, $\delta_{\rm RMS}\theta^{\rm CR2}_{\rm def-MTX} = 14.7~\SI{}{\micro\radian}$, and of the MCS blow-up, $\delta_{\rm RMS}\theta^{\rm CR2}_{\rm def-MCS} = 11.9~\SI{}{\micro\radian}$, induced by the RP0, which is constantly kept into the single-channeled beam to monitor its flux during the data taking. The value of $\delta_{\rm RMS}\theta^{\rm CR2}_{\rm def}$ is, indeed, considerably larger than the critical angle for channeling of Crystal2, $\theta^{\rm CR2}_{\rm c} = 12.25~\SI{}{\micro\radian}$, given in Table~\ref{crystals}. The beam divergence can be reduced by retracting the RP0 in the garage position, however, the mismatch of the beam divergence with the crystal acceptance, due to the accelerator optics, will continue contributing to the reduction of the channeling efficiency and to the increase of the fraction of particles crossing Crystal2 in AM orientation.  In a possible experimental application in LHC, optical configurations should be implemented to avoid this mismatch, obviously detrimental for the physics reach of a double crystal scenario. 

\section{High-statistics beam measurements}

High-statistics runs are performed with the crystals at fixed orientation to measure channeling and volume reflection efficiencies and collect large samples of data in view of disentangling signal from background. In such cases, the alignment procedure requires increased care.

\begin{figure}[t]
    \centering\includegraphics[width=1\linewidth]{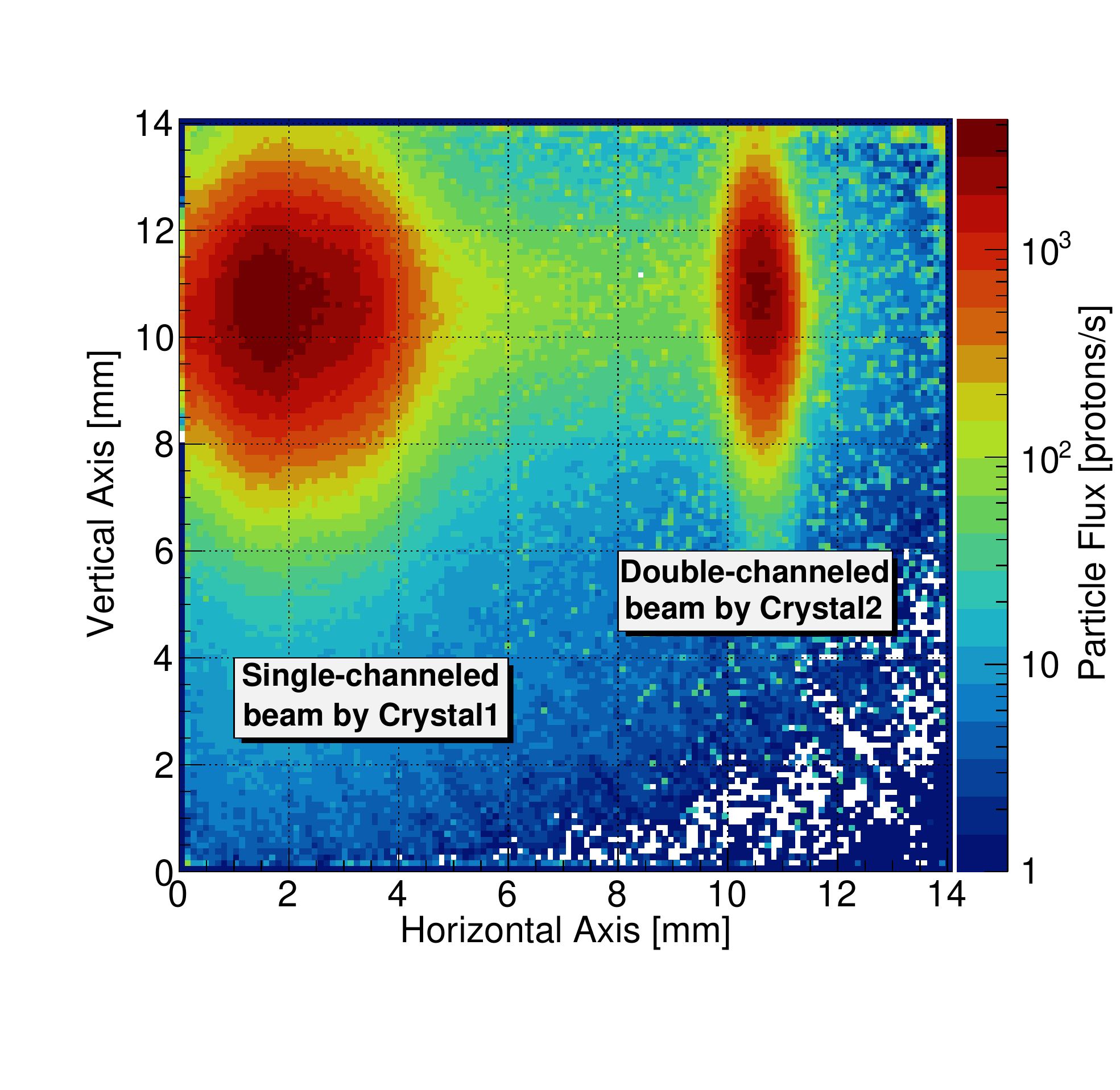}
    \caption{Timepix image of the single- and double-channeled beams. Crystal2 orientation is 1800~\SI{}{\micro\radian}. The right spot is produced by the double-deflected particles. The left spot represents the  particles deflected  by Crystal1 that in the interaction with Crystal2 have not been captured in channeling states. The tiny green cloud between the left and the right spots is due to particles dechanneled during the Crystal2 traversal. The color scale represents the particle flux per pixel. The frame was recorded during a high-statistics run.\\}
\label{fig:fig8}
\end{figure}

\begin{figure}[t]
    \centering\includegraphics[width=1\linewidth]{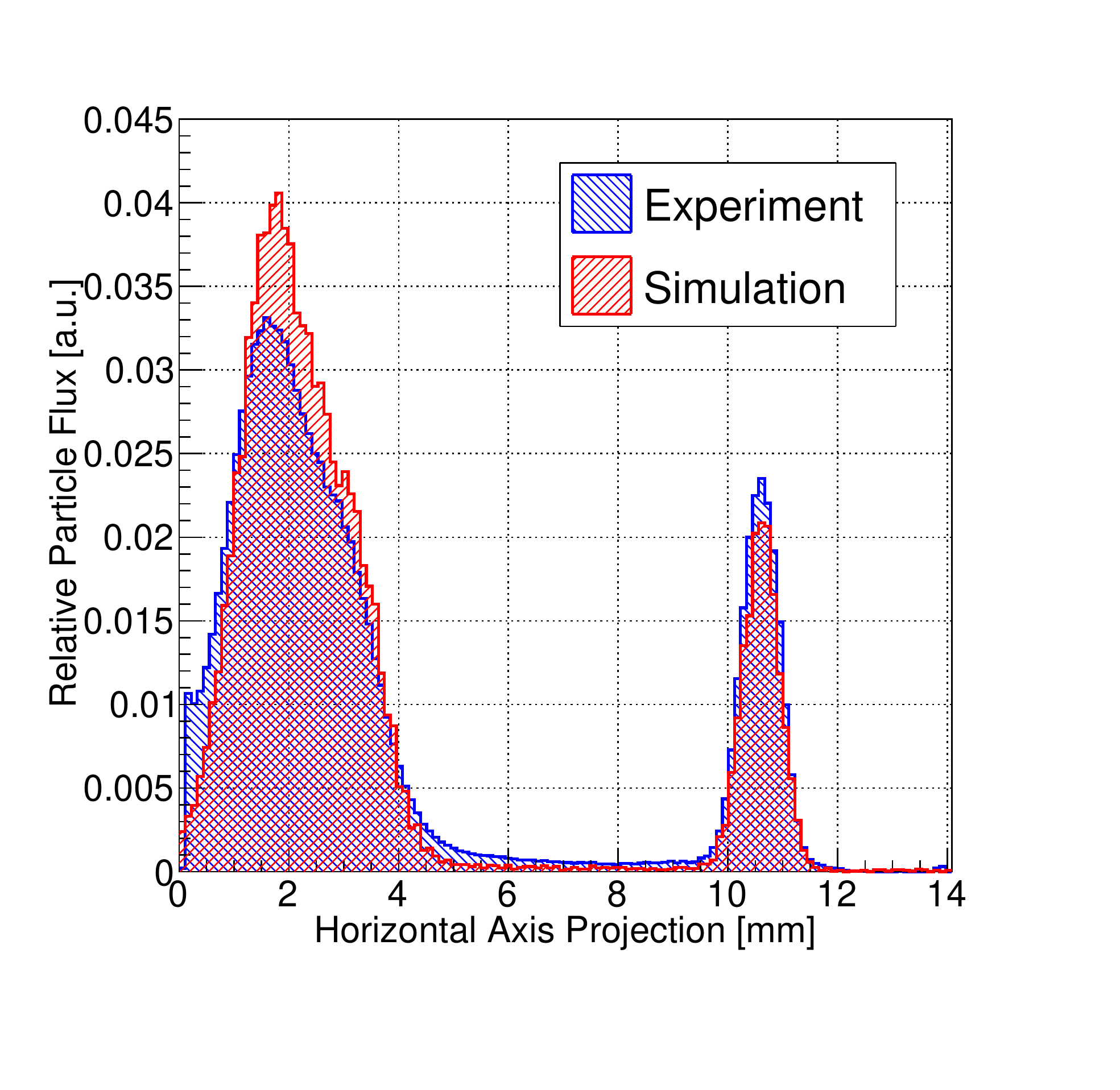}
    \caption{Horizontal profile of  the single- and double-channeled beamlets. The experimental data (blue) are extracted from the frame shown in Figure~\ref{fig:fig8}. In the computer simulations (red), deviations from the optimal channeling orientation  of $35~\SI{}{\micro\radian}$ for Crystal1 and   $20~\SI{}{\micro\radian}$ for Crystal2 are considered.\\}
\label{fig:fig9}
\end{figure}

Figure~\ref{fig:fig8} shows the 2D image of the single- and double-deflected beams, recorded by the Timepix in RP1, in a high-statistics run, with the two crystals in static orientation supposed to be optimal for channeling. The color scale gives the particle flux.
The right spot is produced by the double-deflected particles. The left one represents the single-deflected particles that have interacted with Crystal2, but have not been captured in channeling states. The tiny cloud observed in between the two spots represents the ensemble of particles dechanneled during the Crystal2 traversal. 
The blue curve in Figure~\ref{fig:fig9} shows the horizontal beam profile computed from the data in Figure~\ref{fig:fig8}.
The asymmetric shape of the left beam spot is caused by the coexistence of two particle populations not in channeling states, slightly shifted from each other. One population is made of particles that had amorphous interactions with the crystal  and the other is made of volume reflected particles. 

An \textit{a-posteriori} inspection of the effective displacement of the crystal actuators following the command given to the motors by the remote-control system was recently performed to investigate the effect of a known weakness in the electronic motor-controller.
The analysis suggests that during the crystal alignment an unwanted angular motor slipping could have happened, although the available data are insufficient to precisely evaluate its amplitude. 
An approximate estimate of the misalignment can be obtained, based on the reconstruction of the deflected beam trajectories in Figure~\ref{fig:fig8}, complemented by computer simulations~\cite{routineYellow,mirarchi2015crystal_rout,rossiPhD,dandreaPhD}. The actual orientations of Crystal1 and Crystal2 during the high-statistics run can be obtained as follows. The distance between the single- and double-deﬂected beam baricenters from the circulating beam peripheral at the azimuth of RP1 are known with an uncertainty of 100~\SI{}{\micro\meter}, which is the uncertainty of the RP1 beam-based positioning. Appropriate transfer matrices allow transforming this distance into a crystal orientation through the coefficients $48~\SI{}{\micro\meter}/\SI{}{\micro\radian}$ for Crystal1 and $33~\SI{}{\micro\meter}/\SI{}{\micro\radian}$ for Crystal2, respectively. 
 The resulting value of the actual orientation shift from the optimal angle, previously recorded during angular scans, is  $35\pm2~\SI{}{\micro\radian}$ for Crystal1 and $20\pm3~\SI{}{\micro\radian}$ for Crystal2, the angular uncertainty being originated mainly by the RP1 position uncertainty. 
The red curve in  Figure~\ref{fig:fig9} shows beam profiles corresponding to such misalignment of the two crystals resulting from numerical simulations. The agreement with the experimental data plotted in blue is acceptable, although not particularly good. Moreover, beam profiles computed for slightly different angular shifts are in worst agreement with the experimental data and confirm the soundness of the estimate of the crystal orientation.

Although the orientation of the two crystals  is not optimal during the high-statistics runs, we could evaluate the deflection efficiency from the analysis of the flux of the deflected halo particles. 
The data in Figure~\ref{fig:fig9} can provide the single-pass channeling efficiency of Crystal2, by calculating the ratio between the number of particles in the double-channeled peak, within $\pm 3 \sigma$, and the total number of particles detected in the entire Timepix frame, the latter being the total number of particles deflected by Crystal1. The resulting channeling efficiency of Crystal2 is $\epsilon^{\rm CR2}_{\rm SP}=0.157\pm0.003$, with the error accounting only for the statistical uncertainty. 

\begin{figure}[t]
    \centering\includegraphics[width=1\linewidth]{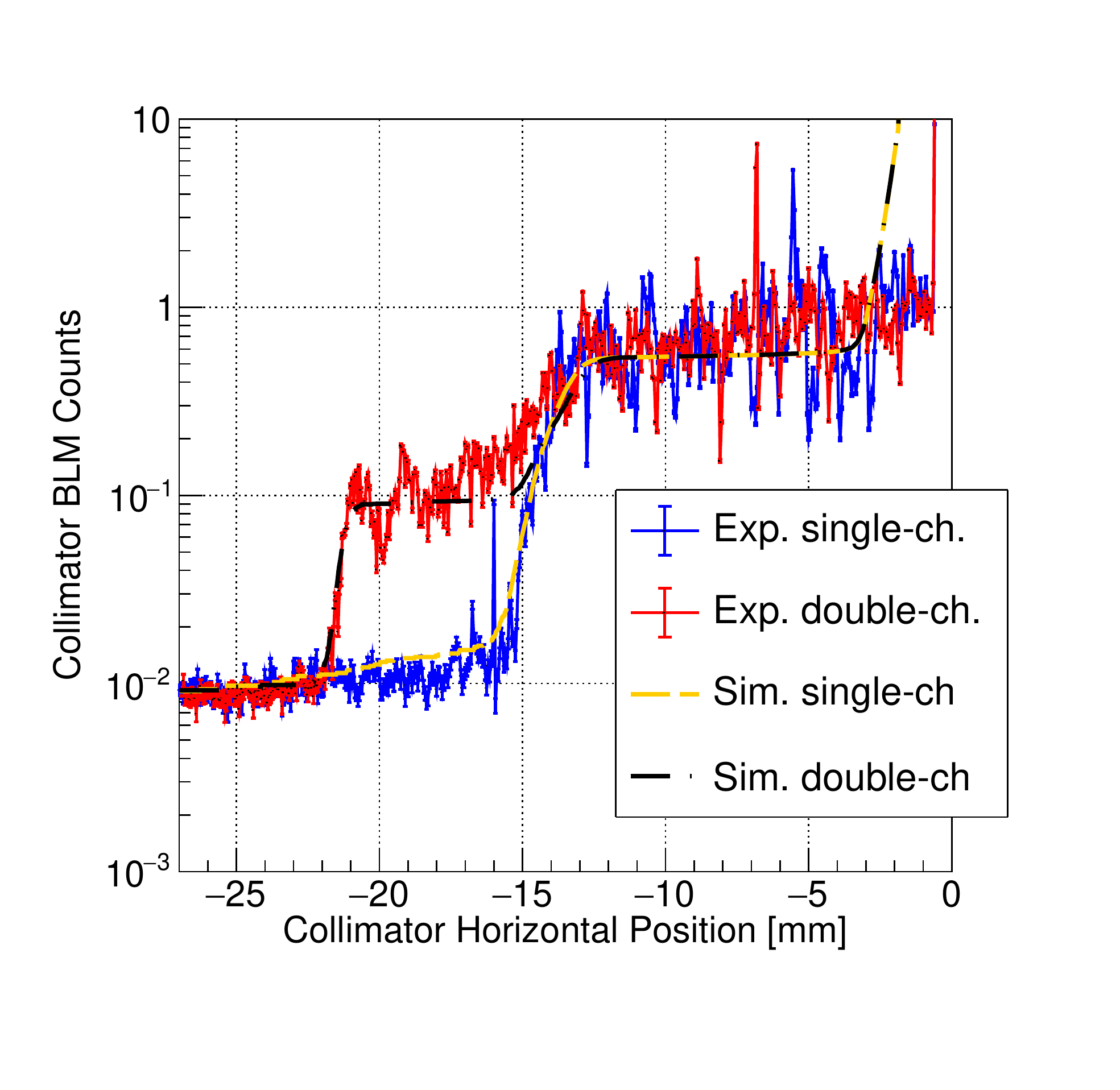}
    \captionof{figure}{BLM signal recorded during linear scans made in single-channeling (blue line) mode and in double-channeling (red line) mode. The orange and black lines show the corresponding simulation results. The BLM counts are normalized to the rate recorded close to the beam edge.\\}
\label{fig:fig10}
\end{figure}

The approach described in~\cite{CHANNELING_22} is used to evaluate the channeling efficiency of the two crystals, using the data of two linear scans recorded at the beginning of the high-statistics run. Figure~\ref{fig:fig10} shows the results of two linear scans performed with one of the collimator jaws swiping across the deflected particles at a speed of 50~$\mu$m/s. The abscissa represents the jaw distance from the circulating beam edge. The ordinate gives the counting rate of the downstream BLM. The blue line shows the case with Crystal1 in channeling orientation and Crystal2 still in AM orientation. The red line shows the case where both crystals are in channeling orientation. Close to the circulating beam edge, the counting rate is proportional to the particle flux impinging onto Crystal1 recorded before the linear scan started, therefore this value is used for normalisation. The counting rate increases while the collimator jaw goes through each deflected beam. After the subtraction of the BLM offset, the channeling efficiency of each crystal is given by the counting rate increase recorded during each beam crossing, i.e from -12~mm to -3~mm for single-channeling and from -20~mm to -15~mm for double-channeling, normalized to the rate recorded close to the beam edge. 

The measured values of the channeling efficiency and the corresponding results from computer simulations are summarized in Tables~\ref{tab:tab3} and~\ref{tab:tab4} and will be discussed in Section 6.
 
 \section{Low-statistics beam measurements}
 
\begin{figure}[t]
    \centering\includegraphics[width=1\linewidth]{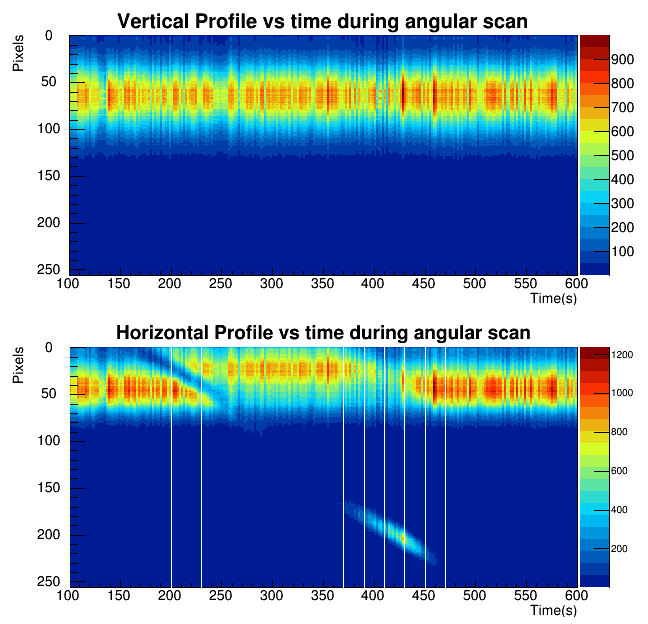}
    \captionof{figure}{Angular scan of Crystal2. Each frame is integrated for 3 s, and each time bin represents the beam projection averaged over 6~\SI{}{\micro\radian}. The nine white lines identify  the time frames used in Figure~\ref{fig:fig12} for  the low-statistics data analysis.    \\}
\label{fig:fig11}
\end{figure}

The data recorded by the Timepix in RP1 during an angular scan of Crystal2 provide similar information as that of Figure~\ref{fig:fig7} for many consecutive orientation angles, but with statistics reduced by about two orders of magnitude. In Figure~\ref{fig:fig11} the vertical and horizontal projections of the recorded RP1 Timepix frames, integrated over 3 s, are plotted as a function of the recording time, during an angular scan of Crystal2 performed at 2~\SI{}{\micro\radian}/s rotational speed. Each time bin represents the beam projection on the relative axis, the vertical in the top plot and the horizontal in the bottom one, integrated over 6~\SI{}{\micro\radian}, i.e over about $0.6\times \theta_{c}$. A color code indicates the particle flux density.
 
In addition to the particles having interacted with the crystal, a continuous flux of background particles, diffusing at high-speed from the circulating beam, is expected to hit the entire Timepix area, with an intensity  decreasing exponentially with the distance from the closed-orbit~\cite{giovannozzi1998dynamic}.
The sectors of the Timepix sensor close to its upper and lower borders are never hit by  the crystal deflected particles all along the angular scan; there, only the diffusive background is recorded. The analysis of the counting rates in these sectors allows identifying  the background shape, which, as expected, decays exponentially with the distance from the center-line and is almost independent of the recording time and of the Crystal2 orientation. 
Spurious hits are also recorded in the vicinity of the sensor border facing the circulating beam; they are due to showers produced by particles traversing the inner side of the RP1 box. In the present experimental configuration, they could not be properly identified and rejected and represent the main source of systematic errors. 
 
\begin{figure}[t]
    \centering\includegraphics[width=1\linewidth]{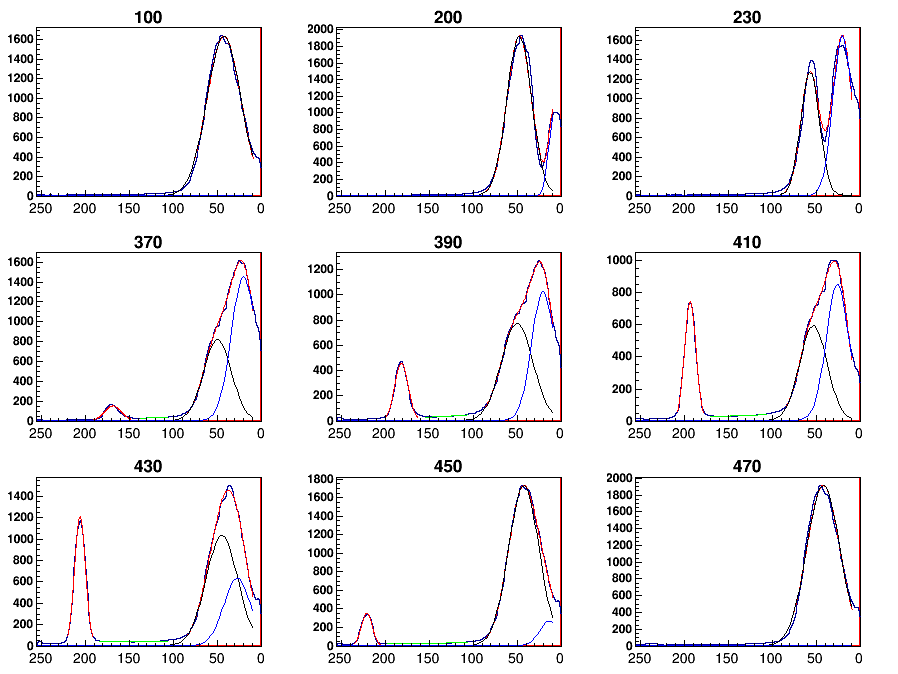}
    \captionof{figure}{Timepix frames collected during the angular scan of Crystal2 shown in Figure~\ref{fig:fig11}. Each frame, integrated over 3 s, shows the number of particle hits per pixel column as a function of the pixel column number. The insert on the top of each plot is the recording time in seconds from the beginning of the Crystal2 rotation. The frames at T=100 s and T=470 s only contain particles in AM orientation. The frames at T=200 s and T=230 s show the onset of VR mixed with AM particles. In the frames from T=370 s to T=450 s, showing the onset and the evolution of the CH process, the left peak contains channeled particles, the right peak in red contains a mix of AM and VR particles, eventually disentangled in the black and the blue Gaussian, respectively. The green curve represents the region of dechanneling. The x-axis gives the distance in mm from the inner side of the Timepix sensor, facing the circulating beam. The y-axis gives the counting rate averaged over 3 s in each pixel column.  \\}
\label{fig:fig12}
\end{figure}

The horizontal profiles corresponding to the beam projections at the nine instants marked by the white vertical lines of Figure~\ref{fig:fig11} are plotted in Figure~\ref{fig:fig12}. Each of the nine data-set is identified by the time elapsed from the start of the angular rotation, shown in the top label, while the abscissa indicates the pixel column number, starting from the inner sensor border, facing the circulating beam, and the ordinate gives the counting rate of the pixel column, integrated over 3 s. The experimental data are plotted in dark blue. They are fitted piecewise to better disentangle the contribution of particles that had different interactions with the crystal. For $x \in[0,107]$, hits of particles having traversed Crystal2 in AM and VR orientations are expected, superimposed to the diffusive background. The fit is therefore made with two Gaussian curves, taking into account the exponential shape of the background intensity. The fitting Gaussian of the VR particles is plotted in light blue, the one of the AM particles in black, respectively, whilst the exponential curve representing the background is not plotted, for the sake of a better readability. The sum of the two Gaussian and of the exponential curves is shown in red. The agreement of the fit with the data is excellent in all the nine plots. For $x \in [107,185]$, no fit is performed: only the exponential background, largely overcoming the expected flux of particles in volume capture, is plotted in green. For $x>185$, only particles deflected in channeling states should hit the sensor plate. The fit is thus made with a single Gaussian, taking into account the exponential background. The sum of the two fitting function, in red, is very close to the shape of the fitting Gaussian of the channeling flux, plotted in light blue, and agrees very well with the experimental data.  The frame 430 corresponds to the optimal alignment for channeling, at an angle of $1820\pm6~\SI{}{\micro\radian}$, where the error takes into account the angular change during the frame integration time.  The dark blue Gaussian to the left side, centered at $x=208$, contains channeled particles; the black Gaussian to the right side, centered at $x=50$, is made of particles in AM orientation; finally the light blue Gaussian, centered at $x=28$, represents particles in VR orientation.  The frames recorded at time T=100 s and T=470 s only contain particles in AM orientation. The frames at T=200 s and T= 230 s show the onset of VR mixed with AM particles. In the frames from T= 370 s to T=450 s, the onset and the evolution of the CH process is shown. The left peak  containing channeled particles is slowly shifted to the left, because the crystal rotation adds-up to the crystal deflection angle, whilst the right peak in light-blue containing the VR particles decreases and the black peak with AM particles increases, because of the continuous shift of the VR tangency point towards the end side of the crystal. It should be pointed out that to enhance the readability of the nine plots, the vertical scale is selected automatically.
 
\begin{figure}[t]
    \centering\includegraphics[width=1\linewidth]{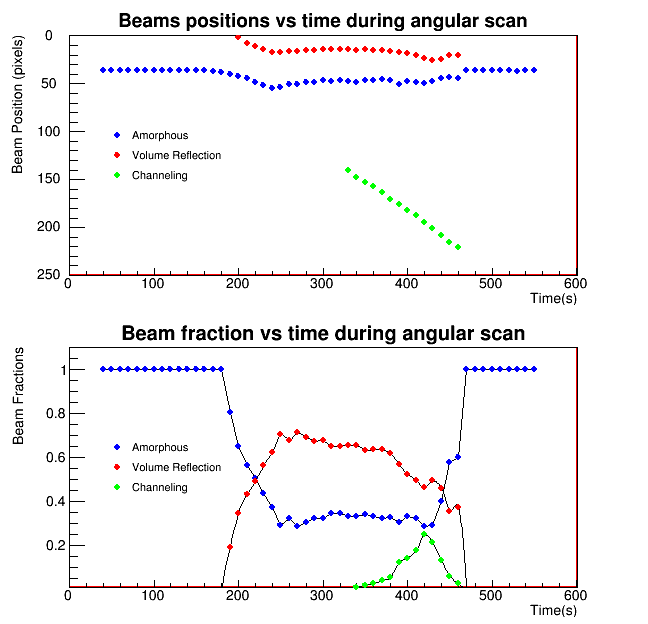}
    \captionof{figure}{Data analysis of the Crystal2 angular scan  of Figure~\ref{fig:fig11}. \textbf{Top:} Time-evolution of the baricenter of the ensemble of particles  interacting with the crystal in AM (blue), VR (red) and channeling (Green) modes, respectively. \textbf{Bottom:} Fraction of particles in AM (blue), VR (red), and channeling (green) states all along the angular scan. The CH fraction is highest at T=430 s.  \\}
\label{fig:fig13}
\end{figure}

Each frame of the angular scan could be used to evaluate the probability of the various crystal-particle interaction types that approximately coincides with the recorded population fraction of each process. For such an evaluation, the exponential background is obviously rejected. Moreover, nuclear interaction events, which cannot be observed in the Timepix sensor, are produced with very small probability, 0.6 \% in AM orientation and 0.03 \% in CH orientation~\cite{scandale2010probability}. They can thus be neglected, without affecting the final results. 
 The data analysis resulting from the entire angular scan is shown in Figure \ref{fig:fig13}, where, in the bottom part, the fractions of particles in AM (blue dots), VR (red dots) and CH (green dots) orientations are plotted as a function of time, whilst, in the top part, the time-evolution of the baricenter of each ensemble of particles is presented.  
 
 In the optimal orientation, at the time frame number 430, the fraction of channeled particles in Crystal2 is $\epsilon^{\rm CR2}_{\rm SP}=0.27\pm0.02$, whilst, during Crystal2 characterisation in the external beam-line H8, it was $\epsilon^{\rm CR2}_{\rm SP}=0.45\pm0.01$ (the efficiency errors only account for limited statistics). The latter value is obtained in H8 with a low divergence beam, selecting the incoming particles within $\pm \frac{1}{2}\theta_c$ angular spread, whilst the former value in double-channeling mode is obtained with a beam divergence of $\pm 1.5 \theta_c$. 
 Indeed, a smaller deflection efficiency is expected in the SPS, because of the contribution of particles incident into Crystal2 at angles larger than $\theta_{c}$, that cannot be captured in channeling states. Assuming a Gaussian distribution of the beam divergence, the deflection efficiency reduction should be of about a factor of two, a value not too far from the ratio between the two experimental values, equal to about 1.7.
 In optimal alignment, with the RP0 inserted, computer simulations predict for Crystal 2 an efficiency of 0.268, perfectly matched to the experiment result provided by the angular scan data. With the RP0 extracted in garage position, instead, the simulation predicts an efficiency of 0.489.
 
 
\section{Summary of the efficiency measurements in the double-crystal configuration}

Measured values and simulation results of the channeling efficiency are summarized in in Tables~\ref{tab:tab3} and~\ref{tab:tab4}. Multi-turn (MT) efficiency
is related to the fact that in a circular accelerator particles can traverse the crystal one or several times before hitting it with the right angle to be channeled. Single-passage (SP) efficiency, instead, refers to the channeling probability for a single crystal traversal. Efficiencies recorded in the SPS, during the double crystal experiment, are due to multi-turn channeling in the case of Crystal1 ($\epsilon^{\rm CR1}_{\rm MT}$) and in the case of the double-channeling process with Crystal1 and Crystal2 ($\epsilon_{\rm DC}$), and are due to single-passage for the Crystal2 alone ($\epsilon^{\rm CR2}_{\rm SP}$). 

Table~\ref{tab:tab3} reports the channeling efficiency from the experimental data and from the corresponding tracking simulations  for the case of the high-statistics run  described in Section 4, with the crystals in fixed orientations, not optimal for channeling,  with misalignments of $35~\SI{}{\micro\radian}$ for Crystal1 and of $20~\SI{}{\micro\radian}$ for Crystal2, respectively, and with the RP0 inserted into the beam.
The first column gives the channeling efficiency resulting from the Timepix data. The second column presents the eﬃciencies based on linear scans and BLM data.  The efficiencies resulting from the  computer tracking simulations based on SixTrack are shown in the third column.

Table~\ref{tab:tab4} shows the channeling efficiency obtained from the Timepix data of the angular scan of Crystal2 described in Section 5, recorded when  Crystal2 is in the optimal orientation for
channeling together with the results of the relative tracking simulations.
The first column gives the efficiency resulting from the angular scan data. For comparison, the second column recalls  the eﬃciency recorded during the Crystal2 characterisation at the H8 beam-line ~\cite{FixedTargetH8}. The third and the forth columns present the efficiencies resulting from  SixTrack simulations with RP0 inserted into the beam and retracted in the garage position, respectively. 

The simulation results and  the results from the experimental data analysis are in rather close and satisfactory agreement, in all cases. The only apparent discrepancy between the measured values of $\epsilon^{\rm CR2}_{\rm SP}$ in columns 1 and 3 of Table~\ref{tab:tab4} has been discussed in Section  5.
Moreover, the simulations show a remarkable  difference in the channeling efficiency in the cases with and without misalignments (third column of Tables~\ref{tab:tab3} and~\ref{tab:tab4}), indicating the margin of deterioration that an inaccurate setting-up of the working configuration can introduce. The alignment control and the matching of the beam divergence with the crystal acceptance for channeling may become of critical importance when proposing a double-crystal setup for higher energy accelerators, such as LHC.

\begin{table*}
\small
\centering
\captionof{table}{Measured deflection efficiencies in the double-crystal configuration, compared with simulation results. The two crystals are in fixed orientations,  with misalignments of $35~\SI{}{\micro\radian}$ for Crystal1 and $20~\SI{}{\micro\radian}$ for Crystal2 from the optimal channeling orientation.}
\label{tab:tab3} 
\begin{tabular}{c|c|c|c}
	\hline
	&Timepix data&Linear scan data&Simulation (RP0 IN)\\
    	\hline 
    	\hline
    $\epsilon^{\rm CR1}_{\rm MT}$&--&$0.519 \pm 0.028$&$0.535 \pm 0.001$\\
    $\epsilon^{\rm CR2}_{\rm SP}$&$0.157 \pm 0.003$&$0.154 \pm 0.014$&$0.151 \pm 0.002$\\
    $\epsilon_{\rm DC}$&--&$0.080 \pm 0.006$&$0.081 \pm 0.001$\\

    \hline
\end{tabular}
\end{table*}

\begin{table*}
\small
\centering
\captionof{table}{Measured deflection efficiencies in the double-crystal configuration, compared with simulation results. The two crystals are in optimal orientation for channeling.}
\label{tab:tab4} 
\begin{tabular}{c|c|c|c|c}
	\hline
	&Timepix data&H8 data&Simulation (RP0 IN)&Simulation (RP0 OUT)\\
    	\hline 
    	\hline
    $\epsilon^{\rm CR1}_{\rm MT}$&--&--&$0.929 \pm 0.001$&$0.929 \pm 0.001$\\
    $\epsilon^{\rm CR2}_{\rm SP}$&$0.27 \pm 0.02$&$0.45 \pm 0.01$&$0.268 \pm 0.001$&$0.489 \pm 0.001$\\
    $\epsilon_{\rm DC}$&--&--&$0.249 \pm 0.001$&$0.454 \pm 0.001$\\

    \hline
\end{tabular}
\end{table*}

\section{Fixed-Target setup measurements}

The effect of a high-Z target was first investigated at the H8 extraction beam-line with a 180~GeV hadron beam~\cite{FixedTargetH8}. During the test in H8, the single-pass deflection efficiency, defined as the ratio of the number of deflected particles to the number of particles entering the complete setup (made either of the crystal alone or of the target+crystal) within an angular range $\pm \theta_{c}$, was measured. The recorded values were $\epsilon^{\rm CR2}_{\rm SP}= 0.45$ and $\epsilon^{\rm CR2+target}_{\rm SP}= 0.11$, respectively. The drastic efficiency reduction on the crystal-target ensemble was induced by the increase of the beam divergence due to MCS into the target and the consequent depletion of the particle population with incident angles within the appropriate angular range for channeling.

\begin{figure}[t]
    \centering\includegraphics[width=1\linewidth]{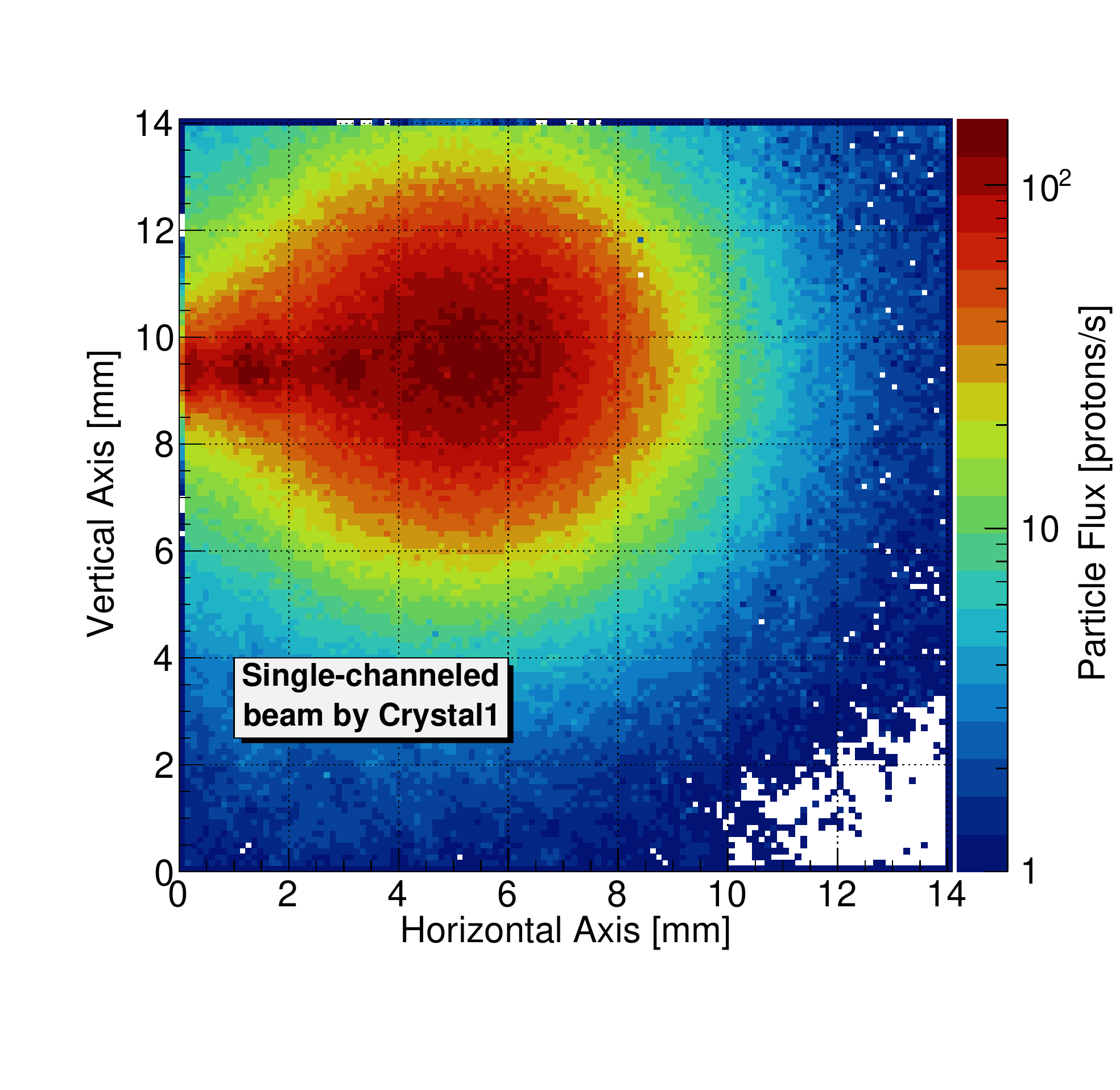}
    \caption{Timepix image of the single-channeled beam after traversing the target and the Crystal2 in AM  orientation, integrated over 0.5 s.\\}
\label{fig:fig14}
\end{figure}

\begin{figure}[t]
    \centering\includegraphics[width=0.8\linewidth]{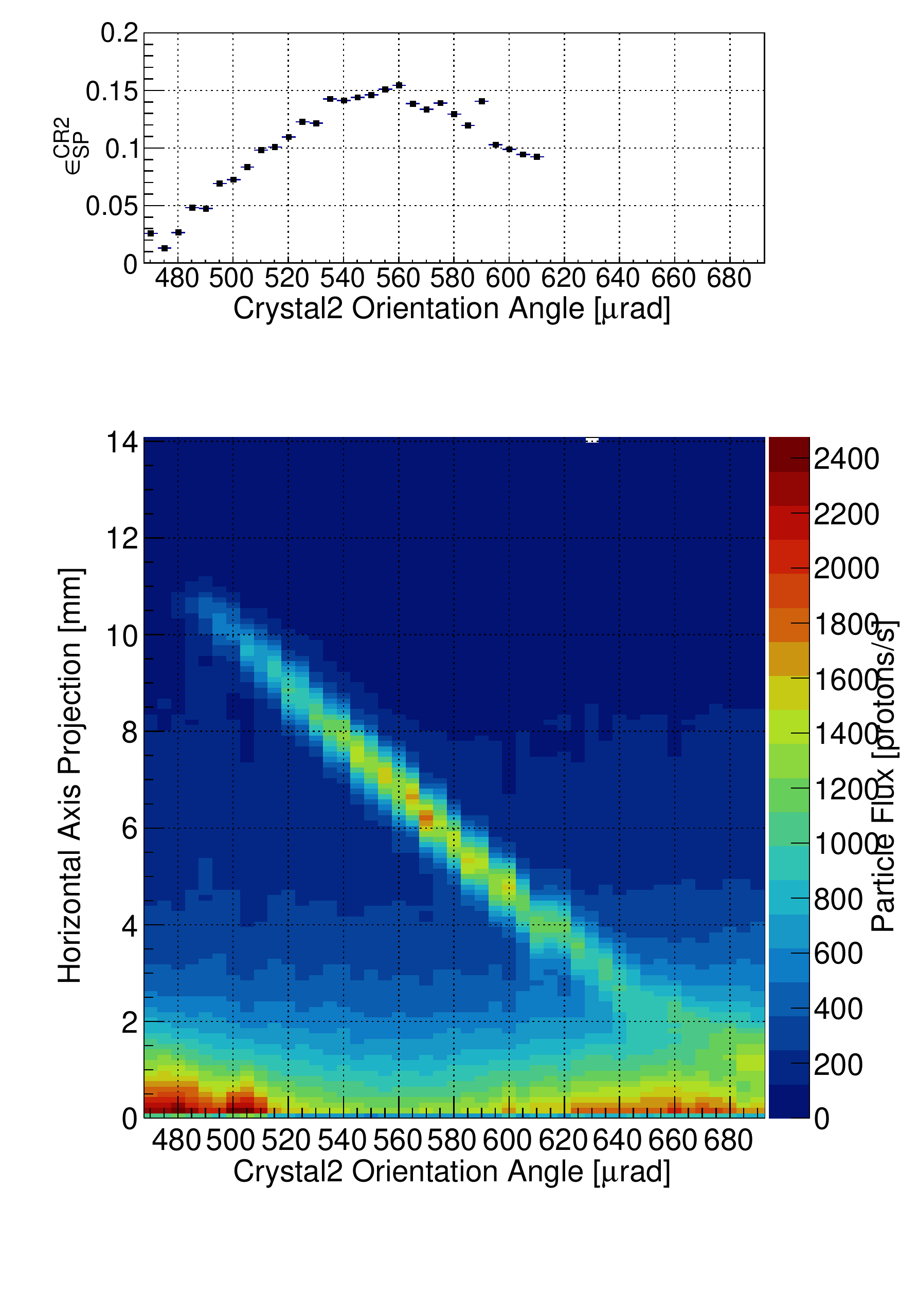}
    \caption{Angular scan of the Target+Crystal2 assembly. \textbf{Top:} Channeling efficiency as a function of the crystal orientation. \textbf{Bottom:} Timepix image (integrated over 0.5 s) of the horizontal projection of the beam intensity as a function of Crystal2 angle. The RP1 is positioned to cover the full range of the double-deflected beam.}
\label{fig:fig15}
\end{figure}

Once tested, the complete setup was installed in the SPS to evaluate the change in performance of the double-crystal scenario. Figure~\ref{fig:fig14} shows the single-channeled proton beam deflected by Crystal1, recorded by the Timepix in RP1, after the traversal of the target+Crystal2 assembly. Crystal1 is in optimal orientation for channeling, whilst Crystal2 is in the amorphous orientation. The measured deflection angle of Crystal1, $\theta_{\rm def} = 301~\SI{}{\micro\radian}$, is equal to the bending angle in Table~\ref{crystals}. Instead, the RMS value of the beam divergence, $\theta^{RMS}_{\rm def} = 46~\SI{}{\micro\radian}$, becomes much larger than $\theta_{\rm c}$, because of the MCS induced by  the target+Crystal2 ensemble. Results from the angular scan of Crystal2 are shown in Figure~\ref{fig:fig15}: in the bottom picture the  projection  of the double-channeled particles on the horizontal axis is displayed, whilst in the top picture the channeling efficiency of Crystal2 as a function of the orientation angle is reported. The angular dependence of the Crystal2 efficiency is strongly modified by the insertion of the target. Indeed, in comparing the two plots of interest, i.e. the green curve in Figure~\ref{fig:fig13} with the curve in the top part of Figure~\ref{fig:fig15}, it becomes clear that the range of crystal orientations for which channeling is allowed is much wider and its peak flatter when the target is added. It is worth clarifying that the origin of angular referential in the two plots is different because of the goniometer calibration, subsequent to the modification of Crystal2.  The bottom view of Figure~\ref{fig:fig15} confirms that the double-channeling spot is visible from 490 to $610~\SI{}{\micro\radian}$. The channeling probability is maximal at $\sim560~\SI{}{\micro\radian}$. In such orientation, the measured single-pass channeling efficiency is $\epsilon^{\rm CR2}_{\rm SP} = 0.154$. The RMS value of the $\epsilon^{\rm CR2}_{\rm SP}$ distribution is $\delta \theta_{RMS} = 0.42~\SI{}{\micro\radian}$ that is, as expected, a value very close to the beam divergence induced by MCS in the target.

\section{Conclusions}

 For the first time, a double-crystal setup was successfully implemented in a circular accelerator. Accurate methods to orient the crystals and record useful data were applied, making use of high-quality crystals, linear and angular actuators, Timepix sensors housed inside Roman Pots and high-sensitivity BLMs located outside of the vacuum pipe.
 Despite the short run time allocated, the experimental procedure was repeated several times, producing deflected beam trajectories and profiles consistently reproducible and close to simulation predictions. 
 
 The optimal crystal orientation was not precisely reached during the high-statistics data-taking run, essentially because of the motor controller instabilities introducing an undesired sliding of the angular actuator. Nevertheless it was possible to reliably measure crystal channeling efficiency from other available sets of data. Indeed, channeling efficiency has proved to be extremely sensitive to even small alignment errors, which could challenge future experiments. In the future an automated online control of the crystal alignment should possibly avoid positioning errors. Moreover, the mismatch of the beam divergence with the angular acceptance for channeling at the entrance of Crystal2 was identified as one of the potential sources of reduction of the channeling efficiency, the other being the increase of the beam divergence during the passage through the target. The run duration was insufficient to investigate the distribution of the induced background and to disentangle the contribution of the crystal from that of the target. This issue, crucial to extrapolate to the LHC the double-crystal scenario, should be addressed in the future.

\section*{Acknowledgements}

The authors gratefully acknowledge financial and technical support from the CERN EN/STI and EN/SMM groups, the UK Science and Technology Facilities Council, the Russian Science Foundation (grant 17-12-01532), and help for the operation of the SPS from BE/OP group.

\bibliographystyle{plane}
\bibliography{ms.bib}

\end{document}